# Angle-resolved photoelectron spectroscopy in a low energy electron microscope

Alexander Neuhaus[1], Pascal Dreher[1], Florian Schütz[2], Helder Marchetto[2], Torsten Franz[2], Frank Meyer zu Heringdorf[1,3]

[1]Faculty of Physics and Center for Nanointegration Duisburg-Essen (CENIDE), University of Duisburg-Essen, 47048 Duisburg, Germany.
[2]ELMITEC Elektronenmikroskopie GmbH, 38678 Clausthal-Zellerfeld, Germany.
[3]Interdisciplinary Center for the Analytics on the Nanoscale (ICAN), 47057 Duisburg, Germany.

## Abstract
Spectroscopic photoemission microscopy is a well-established method to investigate the electronic structure of surfaces. In modern photoemission microscopes the electron optics allows imaging of the image plane, momentum plane, or dispersive plane, depending on the lens setting. Furthermore, apertures allow filtering of energy-, real-, and momentum space. Here, we describe how a standard spectroscopic and low energy electron microscope can be equipped with an additional slit at the entrance of the already present hemispherical analyzer to enable an angle- and energy resolved photoemission mode with micrometer spatial selectivity. We apply a photogrammetric calibration to correct for image distortions of the projective system behind the analyzer and present spectra recorded on Au(111) as a benchmark. Our approach makes data acquisition in energy-momentum space more efficient, which is a necessity for laser-based pump-probe photoemission microscopy with femtosecond time resolution.

## 1. Introduction

The electronic structure of surfaces is intricately connected to their structural composition and is caused by the arrangement of atoms either in simple lattices or superstructures. For instance, in the case of charge density waves the periodic modulation of atomic positions leads to band splitting, highlighting the necessity of combining structural information with electronic characterization for a comprehensive understanding of the system. In heterogeneous surfaces, like domains of 2D materials, this is also evident: rotating graphene domains on a crystallographic substrate (or other graphene layers) can result in metal-semiconductor transitions [1]. Further examples of the interplay between electronic structure and structural properties include two-dimensional materials [2, 3], perovskites [4, 5], vertical heterostructures [6, 7], and microscopically patterned materials [8]. To fully characterize such systems demands a combination of structural analysis and electronic characterization on the nanoscale. Unfortunately, not many methods exist that provide local structural information while at the same time being able to access electronic band structure information with high spatial resolution.

Low energy electron microscopy (LEEM) has been established as an excellent tool to investigate the structural properties of surfaces [9]. In combination with suitable light sources any LEEM can be operated as a photoemission electron microscope (PEEM). Furthermore, imaging analyzers add the possibility to investigate the energy-dependent band structure of selected areas [10]. Recently, LEEMs and PEEMs have been combined with ultrafast laser systems to enable time-resolved photoemission studies of dynamical electron processes in real-space [11] and especially time-resolved studies of optical near-fields [12-14].

Depending on the LEEM or PEEM manufacturer, different sections through the reciprocal space can be recorded: in the SPECS FE-LEEM/PEEM P90 the photoelectron yield $Y(E, k)$ along constant momentum direction $k$ and energy $E$ is directly imaged on the screen (energy-momentum maps). Different directions of $k$ are accessible by rotating the sample [15]. This mode of operation is similar to the one of a modern angle-resolved photoelectron spectrometer (ARPES). In contrast, the analyzer in the ELMITEC SPE-LEEM design provides operation as a momentum microscope, where constant energy cuts $Y(k_x, k_y)$ are recorded as function of a selected energy [10].

Both approaches to record spectral information have advantages and disadvantages that also depend on the research application. For imaging of the surface density of states in the vicinity of the Fermi-level, the operation as momentum microscope [16] is the better option, as direct imaging of cuts through momentum space is possible [17]. In contrast, for recording band dispersions the $Y(E, k)$ mode is the more efficient technique. Certainly, a large volume of the reciprocal space can be accessed using time-of-flight (ToF) photoemission microscopes (PEEM) [18, 19], but time-of-flight techniques are limited to pulsed light sources. When the photoelectron current is low, like in photoemission experiments with low flux light sources or ultrafast laser systems, efficient data acquisition becomes crucial, which also affects which of the acquisition modes is better suited.

Here, we discuss a recent modification to an ELMITEC SPE-LEEM III that provides us – depending on aperture and lens settings – with the option to switch between recording of energy-momentum maps $Y(E, k)$ or momentum microscopy $Y(k_x, k_y)$. Furthermore, we describe a procedure that can be used to correct remaining distortions of the energy and momentum information. We use the well-known band structure of cesiated Au(111) as an example to demonstrate the new mode of operation and the distortion correction.

## 2. Methods

The experiments were performed in the SPE-LEEM III at the University of Duisburg-Essen. The microscope is a standard LEEM instrument (ELMITEC Elektronenmikroskopie GmbH) equipped with a hemispherical imaging analyzer. As described in detail by Schmidt et al. [10], the microscope can be operated in LEEM mode, in LEED mode, and three different spectroscopy modes, such as (i) energy filtered imaging, (ii) energy filtered momentum microscopy and (iii) imaging of the dispersive plane of the analyzer. In the latter, a distinct point in the surface Brillouin zone is selected with a hole aperture (contrast aperture, CAP), the electron distribution is energetically dispersed in the hemisphere, and the projection system of the microscope is used to image the energy distribution of the electrons at the exit of the analyzer on the imaging detector. We improved the electron-optical setup to additionally enable an ARPES-like mode as illustrated in Fig. 1a by introducing an additional 5 µm wide slit (labeled 'Entrance Slit' in Fig. 1) between the retarding lens (RL) and the first inner lens (L1) near the entrance plane of the analyzer. During installation, the long axis of the slit was mechanically aligned to be orthogonal to the dispersive direction of the hemisphere. Since both RL and L1 are electrostatic lenses and do not introduce image rotations, this orthogonal mounting is independent of the settings for RL and L1.

In Fig. 1 beam paths for different electron energies are marked by different colors. Starting at the sample (lower left in Fig. 1a) the focal length settings of the lenses correspond to the standard imaging mode of the SPE-LEEM [10]. The electrons are accelerated from the sample towards the objective lens (Obj), pass through the beam splitter (Sec) and enter into a quadruplet of lenses that provides an almost independent positioning of image and momentum planes. In the standard mode of operation intended by the manufacturer, the plane of the newly installed entrance slit contains an intermediate image plane. Decreasing the focal length of P1 shifts the momentum plane into the entrance slit of the analyzer. Fine adjustment of the momentum distribution relative to the slit position is accomplished with a deflector before RL. The slit cuts out a section of the momentum distribution (Fig. 1b, bottom) that is then dispersed by the hemispherical analyzer (Fig. 1b, top), like it is done in standard ARPES [20], and not unlike it is accomplished in the design based on the dispersion of the beam splitter by Tromp et al. [15]. The slit is imaged into the entrance plane of the analyzer by L1, and a conjugated momentum plane is formed at the exit of the energy analyzer. By decreasing the focal length of the second inner lens (L2) we shift this plane into the plane of the analyzer's exit slit. This plane is then imaged onto a fiber-coupled CMOS detector [21] by the double-lens projective consisting of P2 and P3. Please note that for this operation it is necessary to independently apply potentials to L1 and L2, which in the standard setup are derived from the same voltage source. Operating the microscope in ARPES mode thus requires a slight modification to the wiring within the high-voltage cage. More specifically, the high-voltage supply normally used for RL is now used to supply the high-voltage for both RL and L1, neglecting the thereby induced slight focal shift in RL. Due to the design of RL as a weak immersion lens the changes in focal length can be easily compensated by a slight variation of the focal length of P1. The lens L2 is supplied by the remaining independent high-voltage supply.

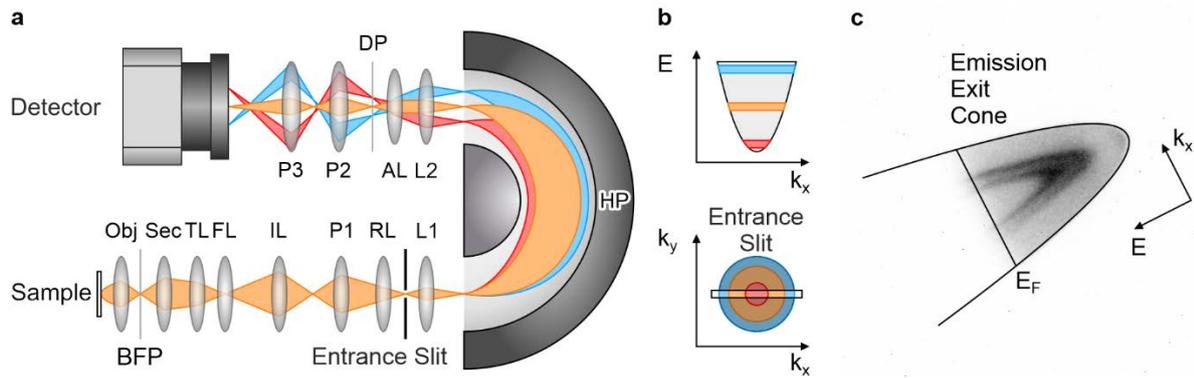

**Figure 1:** Principle of a SPELEEM III operating in ARPES mode. **a** schematic sketch of the imaging lens system in ARPES mode. The electron-optical path for different kinetic energies is shown in blue, orange, and red, respectively. **b** electron distribution in the entrance slit and the dispersed electron distibution in the DP for the three differenct kinetic energies. The entrance slit cuts out a slice of the momentum distribution along $k_x$. The electron distribution is dispersed along the y direction forming an energy-momentum map in the DP. **c** raw data obtained in ARPES mode magnifed by a factor of 4. The electron emisson exit cone and the Fermi edge are shown as guides to the eye. The energy dispersive and momentum direction are indicated by arrows. Two parabolic features appear in the ARPES signal, where the inner one can be attributed to the SS and the outer one can be attributed to direct sp-sp transititons.

The performance of the new ARPES mode is tested using the surface band structure of thermolytically grown self-organized Au(111)-micro-platelets [22, 23] on a Si substrate. After load-locking the platelets were further prepared by several cycles of Argon ion sputtering and annealing until the Au(111) $(22 \times \sqrt{3})$ "Herringbone" reconstruction [24] was visible in LEED and atomic steps were observed in LEEM. Furthermore, to enable linear photoemission using a deep UV-LED ($\lambda$=267nm), a sub-monolayer amount of Cs was deposited to lower the sample's work-function [25].

## 3. Results and Discussion

A typical energy-momentum map that is acquired in the new ARPES mode is shown in Fig. 1c. Micrometer-scale spatial resolution was ensured by inserting a 50 µm wide selective area aperture (SAP) into the beam splitter. More specifically, the aperture is located in a conjugate image plane and selects a circular area of 2.5 µm diameter on the sample. Since only a part of the detector is illuminated by the ARPES signal, the image in Fig. 1c was cropped from the live image the operator sees, and only the central 512 × 512 pixel of the original image are shown. The seemingly arbitrary rotation of the energy-momentum map in Fig. 1c is caused by image rotations in the projective system (P2 and P3) and the orientation of the detector. The approximate energy- and momentum direction of the electron distribution are indicated.

In the ARPES signal the free electron cutoff of the emission exit cone appears as a parabolic edge, and the high-energy end of the electron distribution is cut off by the Fermi edge of the sample. The ARPES signal shows two parabolic features that correspond to the well-known Au(111) Shockley surface state (SS) and a direct sp-sp bulk band transition [26].

## 3.1 Characterization of the distortion field

The magnification of the projective system is comparably large which induces local distortions of the energy and momentum axes in a non-linear manner. Although these distortions are small, and almost not visible in Fig. 1c, a calibration procedure is desirable to correct them and to obtain quantitative energy and momentum information.

For a calibration of the energy and momentum axes the image transfer function of the projective system must be known, i.e., the functional mapping from energy and momentum coordinates in the energy-dispersive plane at the exit of the analyzer onto the detector coordinate system needs to be determined. A common approach to perform such a photogrammetric calibration in an electron microscope is to sample the image transfer function by recording multiple images of the same object that is shifted in a controlled manner [27]. For recording of one-dimensional electron spectra in the dispersive plane, the energy axis is usually calibrated by scanning the sample potential STV relative to the analyzer potential and by recording the shift of the observed electron spectrum on the detector [10]. We expand this procedure to the new ARPES mode to accommodate for the two-dimensional nature of the ARPES spectra.

Figure 2 illustrates the procedure used for energy and momentum calibration. Panels a and b of Fig. 2 show enlarged and rotated ARPES spectra at STV = -1 V and 0 V, respectively. The aforementioned parabolic spectral features are clearly visible. Due to the increase in STV from Fig. 2a to Fig. 2b the ARPES pattern as a whole appears shifted towards the right (lower energy). Closer inspection shows that the marked points $\vec{r}_i$ in the spectra in Fig. 2a are shifted slightly differently from their original detector coordinate to their new coordinate $\vec{r}_i + \Delta\vec{r}_i$ in Fig. 2b. A large number of spectra like in Fig. 2a,b is recorded at monotonically increasing STV and the shift of each point in the spectra is tracked as a function of STV using an optical flow algorithm provided by the open source *Python*-package *scikit-image* [28, 29] .The optical flow is calculated for consecutive image pairs in the STV series and produces a two-dimensional displacement vector field that can be used to transform one image into the other. Note that the displacement field must be independent of STV since it is a property of the projective system and not of the electron spectra. Also, obviously the displacement vectors can only be determined for detector regions that contain signal. Therefore, we mask for pixels outside the region of interest and average the remaining displacement vectors over the STV-series. As a result, we obtain the displacement vector field $\Delta\vec{r}(\vec{r})$ as a function of detector coordinate $\vec{r}$. Figure 2c shows the ARPES signal averaged over the STV series superposed with selected displacement vectors that illustrate which paths particular energy-momentum pixels would take if the STV was increased. Note that for better visibility only few (enlarged) displacement vectors are shown while the full displacement vector field has the same resolution as the detector.

If the image transfer function of the projective system was linear, the energy and momentum calibration would be complete at this point. However, the displacement vectors in Fig. 2c follow curved paths caused by the aforementioned distortions induced by the projective system. Luckily, the displacement vectors can be used to fit a numerical model for the image transfer function of the projective system to the displacement field. The dominant distortions in the magnetic projection lens system are radial and spiral distortions, which can be described by a transformation matrix $\Phi(\rho)$, with $\vec{\rho} = \vec{r} - \vec{r_0}$ for the distance vector between a point $\vec{r}$ and the optical axis of the lens at $\vec{r_0}$ [27]. With this transformation matrix $\Phi(\rho)$ the energy and momentum coordinates $(E, k)$ are related to the detector coordinate system by the nonlinear mapping

$$\begin{pmatrix} E \\ k \end{pmatrix} = \Phi(\rho)\,\vec{\rho}. \tag{1}$$

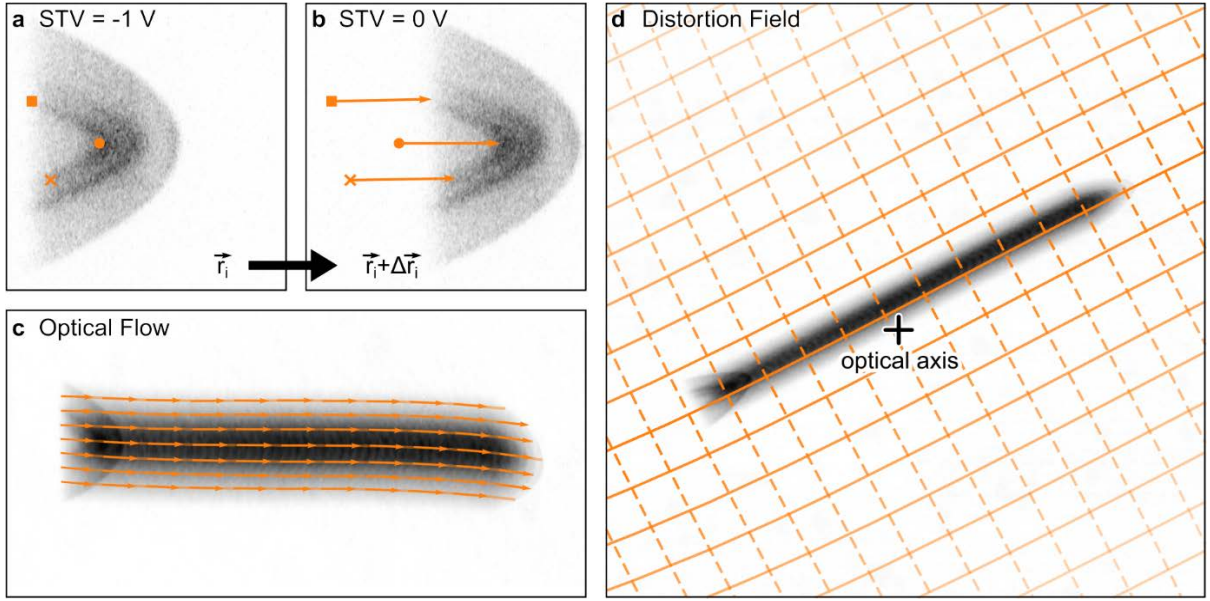

**Figure 2:** Distortion field calibration: **a** and **b** show raw ARPES signals for different sample potentials STV. The ARPES signal is shifted along the energy axis and slightly distorted. The shift of three characteristic points $\vec{r_i}$ is indicated by their shift vector $\Delta\vec{r_i}$. **c** shows the optical flow field (arrows enlarged by a factor of 4) superposing the sum of all ARPES signals. **d** depicts the reconstructed distortion field and the sum of all ARPES signals as seen on the detector. The dashed lines show contours of constant energy, and the solid lines show contours of constant momentum. The extracted optical axis is marked by a plus.

The entries of the transformation matrix

$$\Phi(\rho) = \begin{pmatrix} R(\rho) & -S(\rho) \\ S(\rho) & R(\rho) \end{pmatrix}$$

are given by the radial and spiral distortions $R(\rho)$ and $S(\rho)$, respectively. These distortions can be expanded into power series that only depend on the distance $\rho$ from the optical axis

$$R(\rho) = \sum_{m=0}^{n} R_m\, \rho^{2m};\; S(\rho) = \sum_{m=1}^{n} S_m\, \rho^m,$$

where $R_m$ and $S_m$ are the expansion coefficients that describe different types of distortions [27].

The distortion model depends in a non-linear manner on the absolute positions $\vec{\rho}$ on the detector and not on the displacement vectors $\Delta\vec{r}(\vec{r})$. For small changes $\Delta$STV the displacement vectors will also be small enabling us to linearize eq. 1 and thus relate the displacement vectors to the STV changes

$$\begin{pmatrix} \Delta\text{STV} \\ 0 \end{pmatrix} = \Phi(\rho)\,\Delta\vec{r} + \left(\frac{\partial \Phi(\rho)}{\partial \rho}\,\vec{\rho}\right)\left(\frac{\vec{\rho}}{\rho} \cdot \Delta\vec{r}\right). \tag{2}$$

Fitting eq. 2 to the displacement vector field from the optical flow results in a complete characterization of the projective system's image distortions. Figure 2d shows a representation of the distortion field by plotting contours for constant energy (dashed lines) and constant momentum (solid lines) alongside the averaged raw spectra. The position of the optical axis, a result from the fit, is close to the center of the detector.

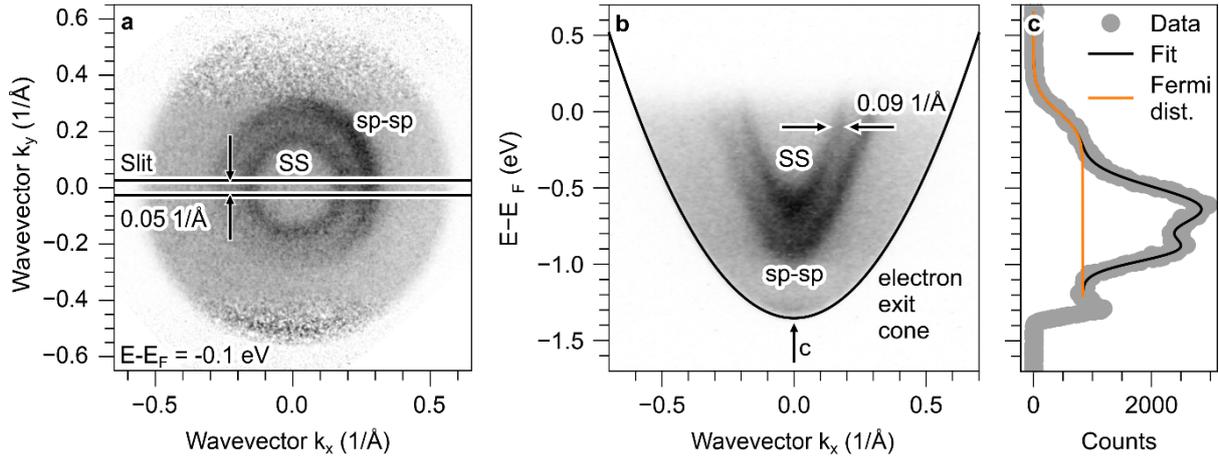

**Figure 3: Spectrum in ARPES mode after calibration**: **a** shows the momentum map 0.1 eV under the Fermi edge. The SS and the direct sp-sp transitions are visible. The width and position of the entrance slit are indicated by a black box. The electron distribution after distortion correction and calibration in ARPES mode is shown in **b**. The Fermi edge is set to zero and the parabolic electron exit cone is drawn in black. The SS and sp-sp transitions yield two more parabolic features in the ARPES signal. A linecut along constant momentum in the vicinity of the $\overline{\Gamma}$-point (gray points), its fit (orange line), and the extracted Fermi distribution (black line) are shown in **c**.

## 3.2 Momentum and energy calibration

For the sampling of the displacement vectors a variation of STV was used and thus the fitted model in Eq. 2 intrinsically provides us with the correct scaling of the energy axis and corrects distortions of the momentum axis. This leaves us with two unknown parameters: the origin of the energy axis, and the origin and proper scaling of the momentum axis. As a contour of constant energy, the Fermi edge provides an intrinsic energy reference and is numerically extracted from the electron distribution.

After proper calibration of the energy axis, the parabolic dispersion of the free electron cutoff of the ARPES spectrum can be used to calibrate the momentum axis and determine its origin. The free electron cutoff is determined by an edge detection filter followed by a generalized parabolic Hough transformation [30].

## 3.3 Resolution

Figure 3 shows data after distortion correction and energy-momentum calibration. From ARPES mode, by decreasing the focal length of P2 and by increasing the focal length of L2 it is possible to image the plane where both the momentum map and the entrance slit of the analyzer are located. This allows us to position the slit and to estimate its width in momentum space. Figure 3a shows an energy-filtered momentum map corresponding to an energy 100 meV below the Fermi energy. The spectral features reflect the surface state and the sp-sp bulk band transitions. The slit width of 5 µm corresponds to a slice in momentum space of width 0.05 Å$^{-1}$ which is then dispersed by the hemispherical analyzer. Figure 3b shows a distortion-corrected energy momentum map with a slit positioned to slice through the $\overline{\Gamma}$-point as marked in Fig. 3a. The effective mass of $m^* = 0.21\, m_e$ of the Shockley surface state is extracted by a parabolic fit to the dispersion of its respective spectral feature. The fitted effective mass is slightly smaller than literature values for pristine Au(111) [31-34], but such a renormalization of the effective mass is known for alkali doping of metal surfaces [35] as is the case for our slightly cesiated Au(111) surface.

In order to obtain an estimate for the attainable energy resolution, a constant momentum cut at the $\bar{\Gamma}$-point is shown in Fig. 3c. The data is fitted with a room-temperature Fermi distribution convolved with a Gaussian function for the instrumental energy broadening and two phenomenological Gaussian functions for the spectral features. The fit yields an overall instrumental energetic broadening of 250 meV. By deconvolution of the 180 meV wide emission spectrum of the used deep UV-LED from the instrumental function of the microscope we estimate an overall energy resolution of the microscope in ARPES-mode of 175 meV. The momentum resolution of the ARPES-mode is not dominated by the microscope, as the momentum width of the electron emission cutoff in Fig. 3b is considerably smaller than the momentum width of 0.09 Å$^{-1}$ of the surface state spectral feature close to the Fermi edge.

## 4. Conclusion

We demonstrated how a standard SPELEEM can be operated as a micro-ARPES spectrometer by inserting an additional entrance slit into the analyzer and changing the lens settings so that momentum information is attainable in the analyzer's dispersive plane. The modification provides a unique combination of structural information accessible by LEEM-based methods with band-structure information accessible by micro-ARPES within the same instrument. More specifically, the photoemission-based possibilites of the SPELEEM are extended from energy-filtered imaging, momentum microscopy, and normal-emission spectroscopy to also provide direct imaging of band dispersion on the screen. We demonstrated a combination of 2.5 µm spatial selectivity with an energy resolution of 175 meV and a momentum resolution clearly sufficient to easily resolve band dispersions after software correction of the distortions of the projective system. The here-shown near-threshold ARPES spectra only cover a small section of the imaging detector, which is unproblematic in a setup with a modern CMOS-based detector [21] with high pixel density and low noise level.

The new ARPES-mode enables surface science studies of dynamical processes that involve the band structure in an SPELEEM, since instead of recording momentum maps at varying sample potential the band dispersion can now be recorded in a single (few seconds) exposure. Here, we only discussed linear photoemission from a Au(111) micro-platelet to demonstrate the technique. However, nonlinear photoemission and pump-probe experiments using femtosecond laser pulses will benefit greatly from the high detection efficiency of the new ARPES mode and the detection of the full emission cone.

## Acknowledgements

This work was funded by the Deutsche Forschungsgemeinschaft (DFG, German Research Foundation) through project B06 of Collaborative Research Center SFB1242 "Non-equilibrium dynamics of condensed matter in the time domain" (Project-ID 278162697). We are thankful to Bettina Frank from Harald Giessen's group at the University of Stuttgart for providing a sample with Au(111) platelets.

## Author Declarations

### Conflict of Interest

The authors have no conflict of interest.

### Author Contributions

A.N., P.D., and F.MzH. developed the idea. F.S., H.M. and T.F designed the slit and installed it. A.N. and P.D. performed the distortion correction and analyzed the data. A.N., P.D. and F.MzH. wrote the manuscript with contributions from all authors.

## Data Availability
The data that support the findings of this study are available from the corresponding author upon reasonable request.